**Freeze casting of hydroxyapatite scaffolds for bone tissue engineering**

Sylvain Deville[1], Eduardo Saiz, Antoni Tomsia

*Materials Sciences Division, Lawrence Berkeley National Laboratory, Berkeley, CA 94720, USA*

[1] Corresponding author: Tel: (510) 486-4817; Fax: (510) 486-4761; E-mail address: SDeville@lbl.gov

Although extensive efforts have been put into the development of porous scaffolds for bone regeneration, with encouraging results, all porous materials have a common limitation: the inherent lack of strength associated with porosity. Hence, the development of porous hydroxyapatite scaffolds has been hindered to non-load bearing applications. We report here how freeze-casting can be applied to synthesize porous hydroxyapatite scaffolds exhibiting unusually high compressive strength, e.g. up to 145 MPa for 47% porosity and 65 MPa for 56% porosity. The materials are characterized by well-defined pore connectivity along with directional and completely open porosity. Various parameters affecting the porosity and compressive strength have been investigated, including initial slurry concentration, freezing rate, and sintering conditions. The implications and potential application as bone substitute are discussed. These results might open the way for hydroxyapatite-based materials designed for load-bearing applications. The biological response of these materials is yet to be tested.

# 1. Introduction

The requirements for a synthetic bone substitute appear deceptively simple, that is, to supply a porous matrix with interconnecting porosity that promotes rapid bone ingrowth, yet possesses sufficient strength to prevent crushing under physiological loads during integration and healing. The ideal bone substitute is not a material that interacts as little as possible with the surrounding tissues, but one that will form a secure bond with the tissues by allowing, and even encouraging new cells to grow and penetrate. One way to achieve this is to use a material that is osteophilic and porous, so that new tissue, and ultimately new bone, can be induced to grow into the pores and help prevent loosening and movement of the implant. Resorbable bone replacements have been developed from inorganic materials that are very similar to the apatite composition of natural bone [1].

In recent years, considerable attention has been given to the development of fabrication methods to prepare porous ceramic scaffolds for osseous tissue regeneration [2-9]. The ideal fabrication technique should produce complex-shaped scaffolds with controlled pore size, shape and orientation in a reliable and economical way. However, all porous materials have a common limitation: the inherent lack of strength associated with porosity. Hence, their application tends to be limited to low-stress locations, such as broken jaws or fractured skulls. Therefore, the unresolved dilemma is how to design and create a scaffold that is both porous and strong.

Freeze casting is a simple technique to produce porous complex-shaped ceramic or polymeric parts [10]. In freeze casting, a ceramic slurry is poured into a mold and then frozen. The frozen solvent acts temporarily as a binder to hold the part together for demolding. Subsequently, the part is subject to freeze-drying to sublimate the solvent under vacuum, avoiding the drying stresses and shrinkage that may lead to cracks and warping during normal drying. After drying, the compacts are sintered in order to fabricate a porous material with improved strength, stiffness and desired porosity. The result is a scaffold with a complex and often anisotropic porous microstructure generated during freezing. By controlling the growth direction of the ice crystals, it is possible to impose a preferential orientation for the porosity in the final material [11].

The technique was applied more specifically to polymeric materials, for tissue engineering. A wide variety of materials have already been investigated, including chitin [12], gelatin [11, 13], collagen [14], PLA [15, 16], PDLLA [15, 17], PLGA [15, 17], poly(HEMA) [18], agarose [19], sericin [20] and alginate [16, 21-24]. Although not stiff and strong enough for load-bearing applications, all these materials have in common a homogeneous structure with open porosity, favorable for rapid cell proliferation. In particular, pore size and its structure can be controlled by heat transfer rate.

We show here how freeze-casting can be applied to hydroxyapatite, an osteophilic ceramic related to the inorganic component of bone, to process bone substitute materials with suitable physical and mechanical properties. In particular, we describe here how the processing conditions (concentration, freezing rate, sintering) affects the scaffold characteristics (size and amount of porosity, compressive strength) and discuss the limits of the technique.

## 2. Experimental techniques

The porous inorganic scaffolds were produced by controlled freezing of hydroxyapatite (HAP) slurries. Slurries were prepared by mixing distilled water with a small amount (typically 1 wt.%) of ammonium polymethacrylate anionic dispersant (Darvan 811, R. T. Vanderbilt Co., Norwalk, CT), an organic binder (1 wt.%, polyvinyl alcohol) and the hydroxyapatite powder in various content (Hydroxyapatite#30, Trans-Tech, Adamstown, MD), depending on the desired total porosity. Slurries were ball-milled for 20 hrs with alumina balls and de-aired by stirring in a vacuum desiccator.

Freezing of the slurries was done by pouring them into a PTFE mold, placed on a Cu cold finger whose temperature is controlled using liquid nitrogen and a ring heater. Typical cooling rates range from 0.5 to 5°C/min. Frozen samples were freeze-dried (Freeze-dryer 8, Labconco, Kansas City, MI) at low temperature (-100 °C, temperature of the cold finger of the freeze-dryer) and low pressure (1300 Pa), for 24 hrs. Sintering of the green bodies was done in an air furnace (1216BL, CM Furnaces Inc., Bloomfield, NJ), in the temperature range of 1250-1350°C, with heating and cooling rates of 2°C/min.

Samples of 4x5x5 mm$^3$ were cut for compression tests. Compression tests were carried out on a testing machine (Instron 1122, Instron, Norwood, MA), with a crosshead speed of 0.2 mm/min. Apparent porosity was derived from the density, measured by Archimede's method. The microstructure of the samples was analyzed by scanning electron microscopy (SEM), and the phases present were identified by X-ray diffraction. SEM work was performed using an environmental scanning electron microscope (S-4300SE/N, Hitachi, Pleasanton, CA). X-ray diffraction was performed using a Siemens D500 diffractometer (Siemens AG, Munich, Germany).

## 3. Results

By directional freezing of the slurry, we force the particles in suspension to be rejected from the moving ice front and piled up between the growing columnar ice crystals [2]. Afterwards, the ice is sublimated by freeze-drying, such that a ceramic scaffold whose microstructure is a negative replica of the ice is produced. The porosity of the sintered materials is a replica of the ice structure before drying.

### *3.1 Sample general features*

According to their microstructure, the samples can clearly be divided into three distinctive zones (Fig. 1), each of them characterized by a different shape and dimensions of the pores. In zone 1, the closest to the initial cold finger, no porosity at all is observed and the material is dense. In the second zone, the material is characterized by a cellular morphology. Finally, in the upper zone (zone 3), the ceramic is lamellar, with long parallel pores aligned in the movement direction of the ice front. The underlying reason for this architecture will be described later in the discussion. Unless explicitly stated otherwise, this paper will focus on the last zone (zone 3).

The microstructure of the lamellar zone is to some extent very similar to that obtained on polymeric materials for tissue engineering processed with the same technique. It consists of hydroxyapatite plates with flat interconnected macropores between them, aligned along the ice growth direction (Fig. 2). On the internal walls of the lamellae, a dendritic, branch-like structure can be observed following the microscopic ice formation. Some ceramic bridges, linking adjacent plates, are also observed. The sintered scaffolds had relative densities between 30 and 70% depending on the solids concentration in the initial slurries and the width of the open interconnected macropores, which typically ranges between 20 and 100 µm in their smallest dimension and 50 to 500 µm in the largest one.

## *3.2 Phase analysis*

The XRD pattern (Fig. 3) of sintered samples can be completely indexed with hydroxyapatite (JCPDS 09-0432), the only phase found present. No processing residue or secondary phases were found in the materials.

In the following sections, we describe the effect of the processing conditions, namely, the slurry concentration, freezing rate, and sintering, on the main properties of interest for tissue engineering: shape and amount of porosity, and compressive strength.

## *3.3 Influence of slurry concentration*

### - Total porosity

By changing the initial slurry concentration, a broad control of the final porosity can be achieved (Fig. 4). The relationship between the final porosity in the lamellar zone and the initial slurry concentration is linear, provided sintering conditions are constant. Not only does the technique offer very reproducible results, but also the final porosity can easily be tuned by the initial slurry concentration.

### - **Porosity size**

For highly porous samples (Fig 5a), the microstructure is almost cellular (bridges and walls have similar dimensions). Walls can be made very thin (<5µm), especially for fast cooling rates (>5°C/min), and their thickness is very homogeneous throughout the whole sample. For lower porosity content (47%, Fig 5b), some scattering is observed in the distribution of wall thickness, as repelling particles from the ice front becomes more difficult with increasing particle content.

**- Compressive strength**

Compressive strength vs. total porosity is plotted in Fig. 6. Although for high porosity content (typically >60 vol%) the strength (16 MPa) is comparable to that reported in the literature, it increases rapidly when the porosity decreases, reaching 65 MPa at 56% porosity and 145 MPa at 47% porosity. Values obtained for these samples are well above those reported so far in the literature. In fact, the strength of the porous lamellar HAP is close to that of compact bone [25]. Such high values allow considering the potential of these materials for some load-bearing biological applications.

## *3.4 Influence of cooling rate*

- **Porosity size**

Being highly anisotropic, the pore channels have been characterized here by two dimensional parameters: the long axis and the short axis. The pore size can be modified mostly by increasing or decreasing the cooling rate of the cold finger during freezing. For low cooling rate (<1°C/min), the final microstructure will be characterized by large lamellae thickness (>50 µm) and pore width (>500 µm in the long axis and 40 in the short axis, Fig. 7). Conversely, for fast cooling rate (>5°C/min), finer lamellae and pores are obtained.

*-* **Compressive strength**

A series of samples with identical slurry concentration (50 wt.%) have been processed using similar conditions, the only difference being the cooling rate during freezing. The total porosity is independent of the cooling rate and was the same for all the samples, as expected. It has previously been shown how the cooling rate strongly affects the porosity features (size and morphology). These changes in cooling rate have, consequently, a direct repercussion on the compressive strength (Fig. 8). For low cooling rate (<2°C/min), the compressive strength remains moderate, at 10 to 20 MPa. However, for fast cooling rates (>5°C/min), the compressive strength increases up to 60 MPa as the cooling rate increases. This effect can be interpreted by considering the effect on the microstructure previously illustrated in Fig. 7. Porosity size (i.e. defect size) decreases as the cooling rate increases, so that the strength also increases as the defects (pores) become smaller. This also explains why the rate of increase in strength diminishes for fast cooling rates (>5°C/min). Indeed, the relationship between porosity size and cooling rate is not linear (Fig. 7), so that it becomes more and more difficult to reduce the pore size by accelerating the freezing. Samples cooled at 7 and 10°C/min exhibit similar porosity dimensions, and therefore have similar compressive strength. The technical limits of the experimental setup do not allow cooling rates greater than 10°C/min, and biological response should be taken into consideration before trying to size down the porosity as much as possible.

### *3.5 Influence of sintering conditions*

The sintering conditions are of prime importance for optimizing the physical and mechanical properties. Porosity and compressive strength were measured as a function of the sintering temperature (sintering of 3 hrs). Results are given in Fig. 9. When sintering temperature increases from 1250°C to 1325°C, porosity goes down and compressive strength goes up. Both then reach a plateau value, which means that densification is achieved at 1325°C. Further increase in sintering temperature will only increase the grain size. Therefore, the optimum sintering temperature was determined to be 1325°C.

It is also worth noting the increase of grain size with temperature is very limited, as compared to that observed with materials processed by conventional techniques (e.g. [40]). This particular behavior of freeze-dried materials was already observed for freeze-dried ceramics [41]. Although the underlying reasons of this behavior are still to be elucidated, it is probably partly related to the pinning of grain boundary at the surface during sintering of thin films, since the grain size lies in the same range of order as the lamellae thickness.

## 4. Discussion

### *4.1 Control of the microstructure and the physics of ice formation*

The final microstructure is a replica of the ice. Modifying the amount of water in the slurry as well as the shape of the ice crystals will modify the final amount, shape and size of porosity of the porous ceramic scaffold. In particular, during the steady freezing regime, the ice crystals exhibit a homogeneous morphology throughout the whole sample; this explains why the lamellae thickness is very homogeneous throughout the whole sample. This also means that the ratio porosity thickness/layer thickness is primarily determined by the initial slurry concentration. For instance, if the volume fraction of ice and ceramic particles is the same, layer thickness (e.g. 20 microns) will be the same as the porosity (20 microns in their smallest dimension). For low slurry concentration, the porosity becomes predominant and layer thickness decreases.

The microstructural features can be controlled by applying the physics of ice formation. In particular, the ice-tip radius (and as a consequence the thickness of the ice crystals), which is physically determined by the magnitude of supercooling ahead of the freezing front [42], can be modified by increasing or decreasing the cooling rate during freezing. For fast cooling rates, supercooling becomes larger and the microstructure can be scaled down. On the other hand, under a very slow cooling regime the layer (or pore) thickness can be noticeably increased (Fig. 7).

## *4.2 Physical limits of the technique*

Fundamentals of the physics of ice will determine the limits of the technique. In particular, the sequence leading to a steady lamellar ice front is well known and involves a progressive transition of the freezing front morphology from planar to cellular and cellular to lamellar (also called dendritic), due to the progressive build up of supercooling ahead of the freezing front. This evolution is replicated in the microstructure close to the initial freezing front, i.e. the one initially in contact with the cold finger, as seen in Fig. 10. If the zone corresponding to the planar to cellular zone (Fig. 1) is difficult to ascertain, the cellular-to-dendritic transition zone is clearly observed in the middle of the micrograph. Once the transition has occurred, a steady regime is reached, and the freezing front retains its morphology. Hence, two very different architectures are obtained in the final part. The lower part, formed during the cellular regime, exhibits a porous structure, but with poor pore interconnectivity. The average pore size is small; from a biomedical point of view, this will be of little interest and this part will probably have to be removed. The upper part exhibits the porous lamellar architecture, with open and interconnected porosity. Although no further experiments were performed at this stage, the size of the transition zone seems highly dependant on cooling rate, slurry concentration and ceramic particle size.

Although already observed in freeze-dried polymeric materials (e.g. see fig. 2a in [12] and fig. 1a in [22]) the presence of such transition zones has not yet been explained. Combining the physics of ice and the interaction of the freezing front and inert particles, the present and previous experimental results can be interpreted in terms of modification of the freezing regime and transition of the freezing front morphology. The first zone in contact with the cold source exhibits faster freezing kinetics, and the equilibrium in terms of supercooling and heat diffusion has not been reached yet. Hence, the freezing interface exhibits an unstable cellular morphology, resulting in poor pore interconnectivity.

The influence of slurry concentration on the final porosity has been illustrated previously. The range in which porosity can be adjusted is physically limited by two phenomena. At low concentrations, even in the presence of a binder, collapse of the green body occurs during the sublimation of the ice. However, above a certain concentration (65wt.%), the interaction between particles in the slurry becomes more important and eventually particles cannot be repelled from the ice front anymore. As a result, the interconnected and open porous structure is lost (Fig. 11). Such architecture is not suitable anymore for tissue engineering.

This critical slurry concentration is strongly dependent on the particle size. Earlier theoretical and experimental results have established the existence of a critical velocity for particles engulfment [43, 44]. If the velocity of the front is above the critical value, particles will be engulfed by the interface. In addition, it was found that the critical velocity is inversely proportional to the particle radius [44]. Repelling small particles from the ice front is easier than larger ones. Hence, by using a powder with a smaller granulometry (the particle size of the powder used here was 2.4 µm, data provided by manufacturer), reaching lower porosity while retaining the interconnected open porosity becomes feasible.

## *4.3 Suitability as bone substitute*

Klawitter and Hulbert [45] established a minimum pore size of ~100 µm for bone growth into ceramic structures, and a similar conclusion was reached by Simske et al.[46]. More recently, Itala claimed that bone ingrowth occurred in pores as small as 50 µm [47], and similar results were obtained for porosity engineered in the range 15-40 µm [48]. The porosity must be interconnected to allow the ingrowth of cells, vascularization and diffusion of nutrients. Typical porosities for HA scaffolds described in the literature range from 35% [35, 46] to 75% [49] with pore sizes between 50 µm [47, 50] and 400 µm [49, 51-53]. A preponderance of pores this large can make the implant material rather weak. Moreover, even small movements of the implant

(which are hard to avoid) can cause complications by cutting off blood supply to tissue in the pores, which can lead to inflammation. In addition, bone response reflects meso-scale structure, and direct bone attachment and conduction along the struts of porous scaffolds has been reported [22] even for small porosity [48], revealing that pore connectivity and orientation are at least as important as pore size. Hence, the porosity range reported in this study might be suitable for osseous regeneration.

An additional feature might be taken into consideration, namely the dendrites surface relief covering the surface of the lamellar walls (Fig. 12). Considerable attention has been paid recently to the relationships between cellular response and morphological features such as mesostructure, patterns, roughness, etc. Particularly, this was made possible with the recent development of computer-assisted processing techniques. Results from a wide range of experimental conditions pointed out the beneficial effects of roughness, patterns and architectures, which seem to shorten the time necessary for new bone ingrowth and tissue formation [19, 54-61]. The presence of such patterns allows directional growth of the cells and faster vascularization of the implants. In our case, these features, originating from the dendritic shape of the growing ice crystals during freezing, are also oriented along the ice growth direction and hence the porosity. By adjusting the slurry concentration and freezing conditions, their size might be controlled to some extent, so as to match more favorable dimensions of guidance patterns [60], improving the osteoconduction and osteoinduction characteristics. Experiments are under way to assess the validity of these hypotheses and the biological response of these materials.

It is finally worth mentioning that, because the process relies on physical and not chemical interactions, it can be easily extended to any calcium phosphate or suitable ceramic for biomedical applications. The unique features of the porosity obtained by this technique have already attracted attention for non-load bearing applications using polymeric materials. The technique can now be extended to load-bearing applications.

# 5. Conclusions

Based on an experimental study of freeze-drying of hydroxyapatite powders with various slurry concentrations and sintering conditions, the following conclusions can be made:

1. Porous scaffolds with total porosity ranging from at least 40% to 65% can be obtained by freezing of hydroxyapatite aqueous suspensions and subsequent ice sublimation and sintering. The resultant porosity is open and unidirectional, exhibiting a lamellar morphology. Size of the porosity can be controlled by modifying the freezing rate of the slurries and the slurry concentration.

2. Formation of the porosity can be understood by applying basic principles of the physics of ice and the interaction of a moving solidifying interface (the freezing front) and inert particles in suspension. In particular, the influence of cooling rate, particle size and slurry concentration can be interpreted.

3. Due to their lamellar architecture and the pore shape anisotropy, the processed scaffolds exhibit unusually high compressive strength for such materials, i.e. up to 145 MPa for 47% porosity and 65 MPa for 56% porosity, allowing them to be considered for some load-bearing applications.

**Acknowledgements**

This work was supported by the National Institute of Health (NIH/NIDCR) under grant No. 5R01 DE015633 (Novel Scaffolds for Tissue Engineering and Bone-Like Composites).

Figures

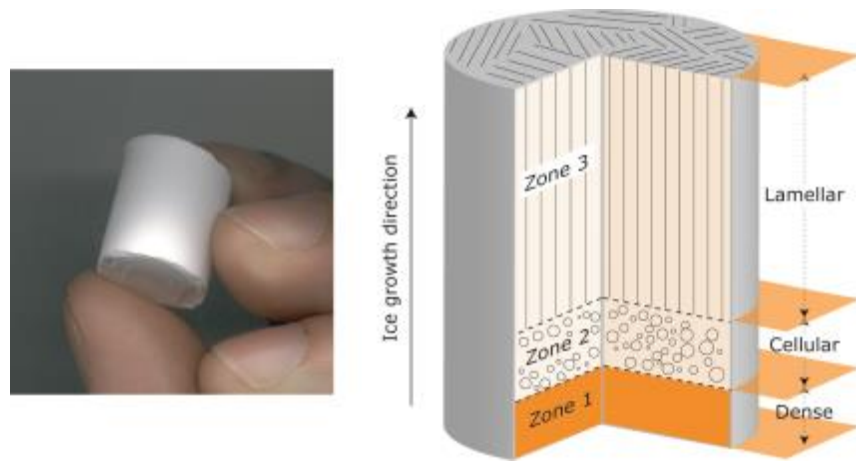

*Fig. 1: Typical sample (left) and general microstructure of the porosity (right). Three distinctive zones are found, their dimensions depending largely upon processing conditions.*

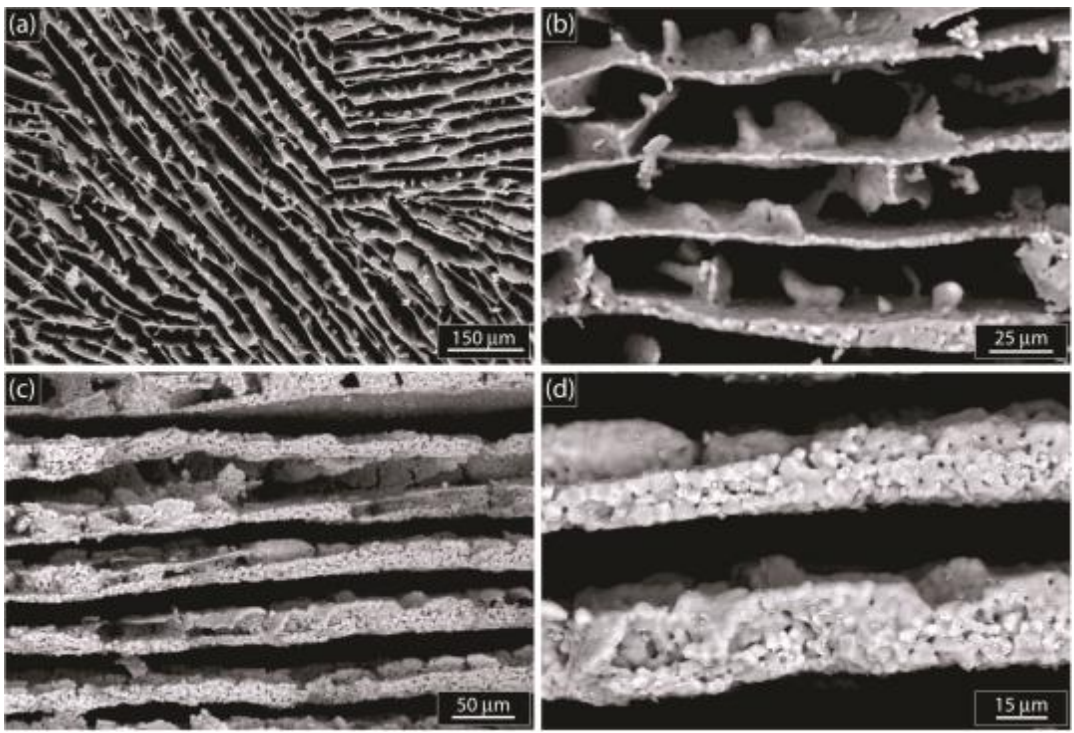

*Fig. 2: Microstructure of porous HA samples with 64% porosity. Cross-sections parallel to the ice front (a) and detail (b), and perpendicular to the ice front (c) and detail (d). The slight difference in wall thickness (b and d) arises from the presence of a small gradient of density at the upper surface of the sample (a and b), due to particle segregation at the top of the mold during freezing.*

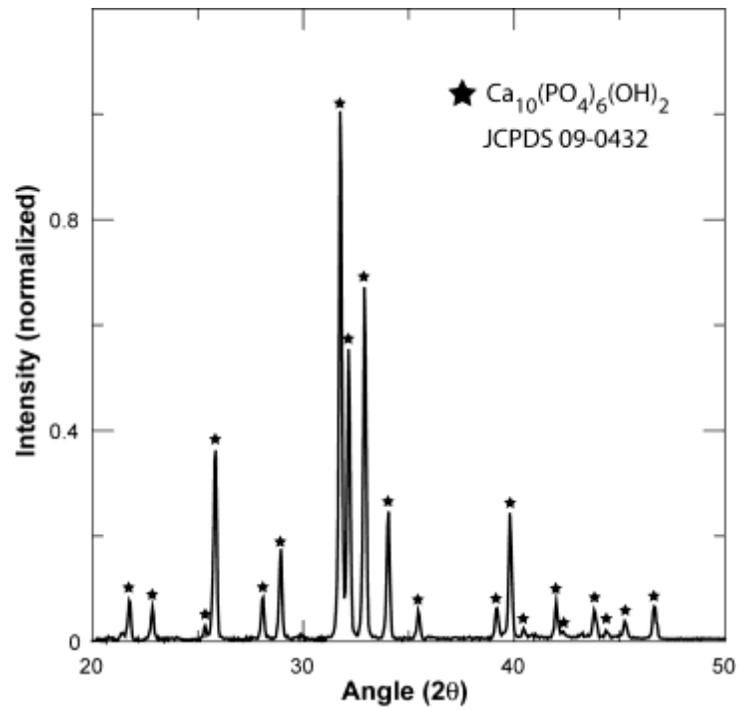

*Fig. 3: XRD spectrum, sintered materials.*

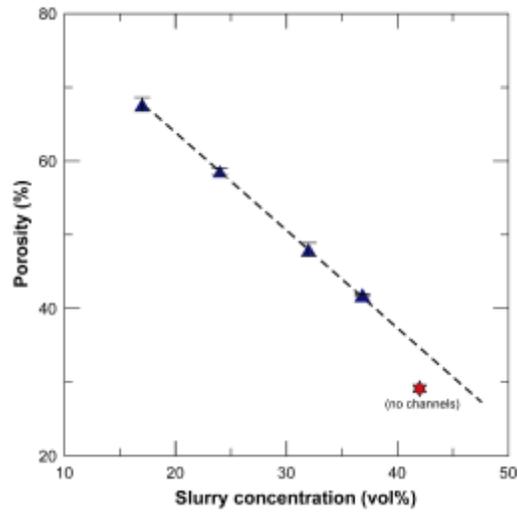

*Fig. 4: Total porosity in sintered materials vs. slurry concentration. The fit is plotted for the lamellar microstructures only.*

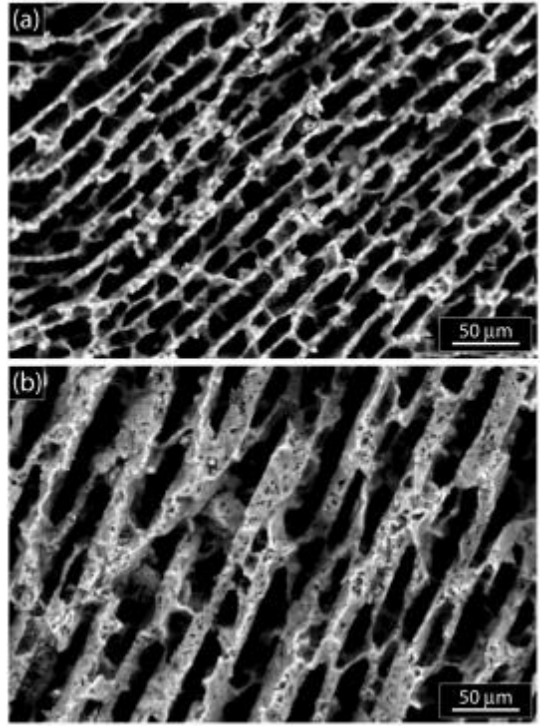

*Fig. 5: Effect of initial slurry concentration. Microstructure of porous HA samples, parallel to the ice front with (a) 64% porosity and (b) 47% porosity.*

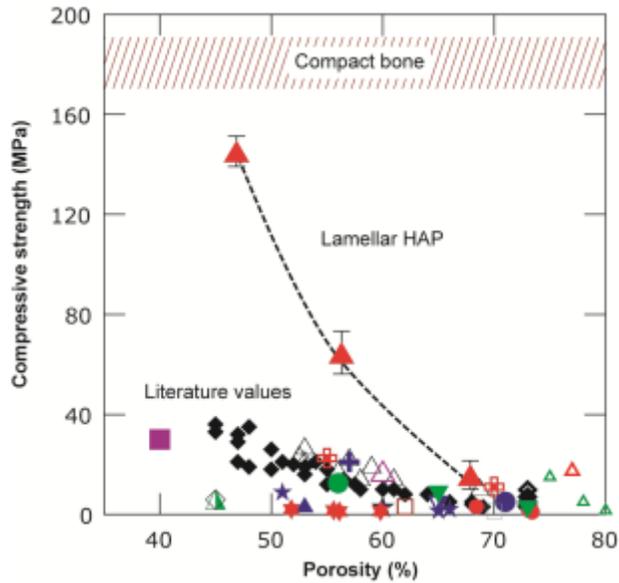

*Fig. 6: Maximum compressive strength achieved by freeze casting (cooling rate of 5°C/min) vs. porosity. Comparison with compact bone [26,27] and literature values [3-9, 28-39] of porous HAP for tissue engineering, conventional and computer-assisted processing techniques. Each style of point corresponds to a different literature source.*

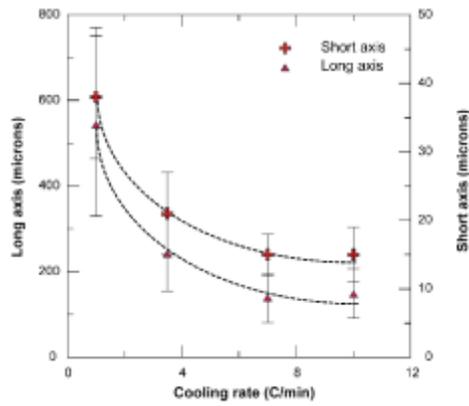
*Fig. 7: Influence of the freezing rate over the pore width (56% total porosity).*

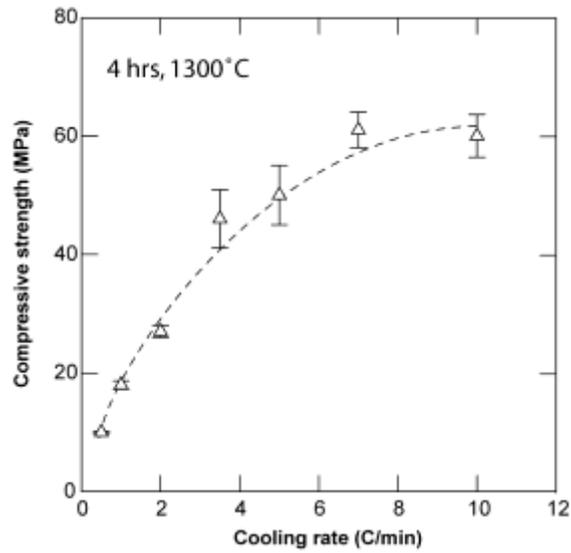
*Fig 8: Compressive strength vs. cooling rate. Samples with 56% porosity. Sintering 4 hrs at 1300°C.*

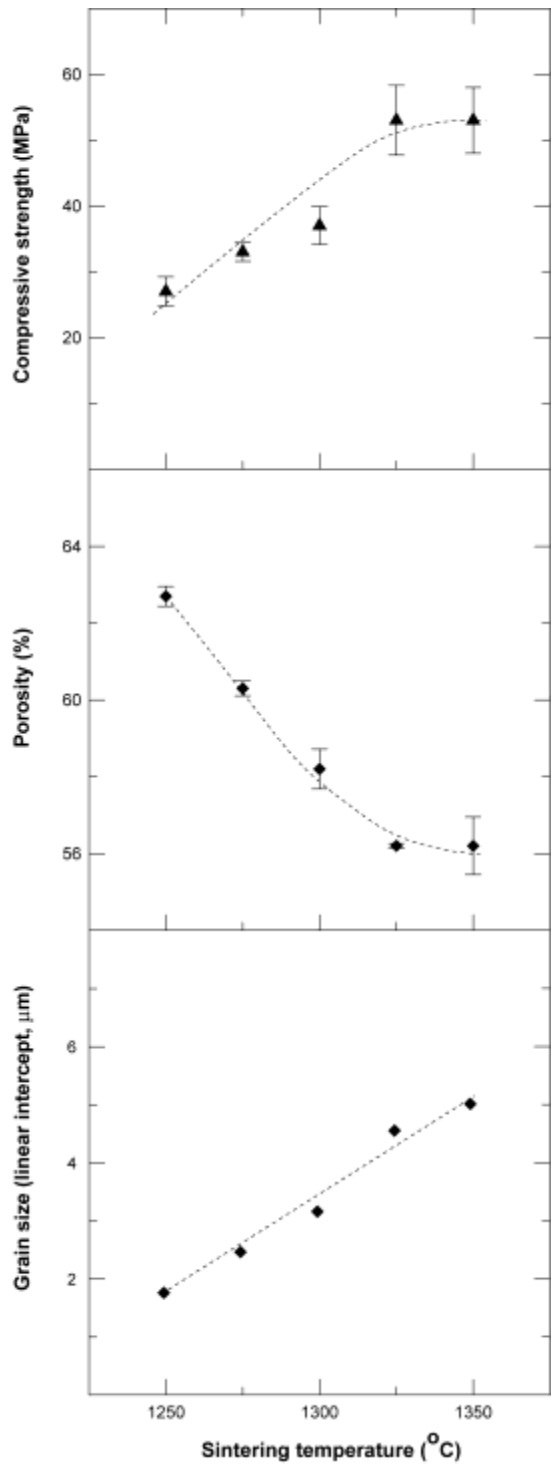

*Fig. 9: Influence of sintering temperature (3 hrs at dwell temperature) on compressive strength, total porosity and grain size. Samples with 24 vol% of initial powder, frozen at 5°C/min.*

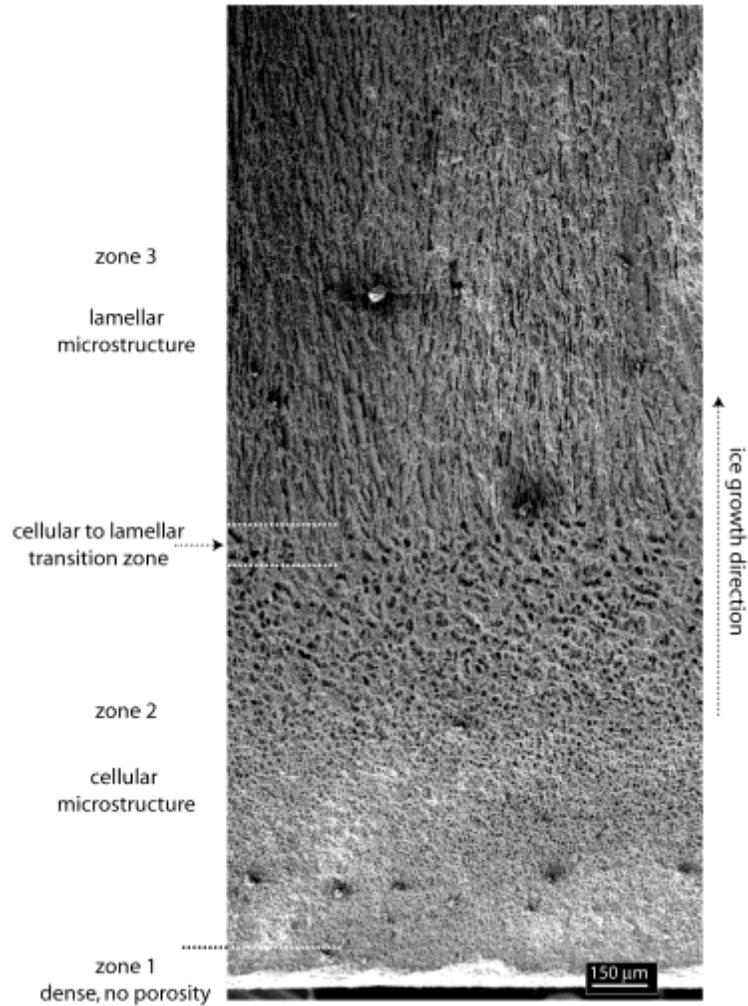

*Fig. 10: Microstructure of porous HA samples, close to the cold finger (cross section perpendicular to the ice front). The initial porosity gradient is related to the evolution of the freezing front morphology with time. The cellular to lamellar transition zone is clearly observed in the middle of the micrograph.*

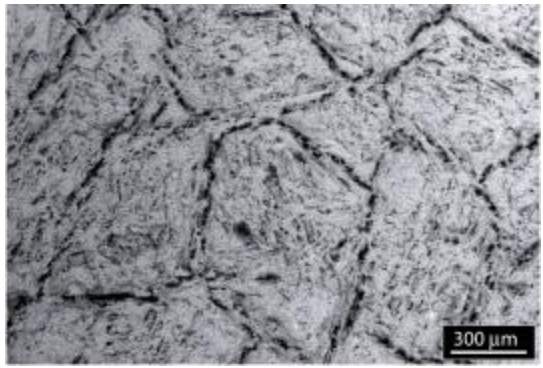

*Fig 11: Cross section parallel to ice front for initial slurry concentration of 42 vol%. The interconnected and open porous structure is completely lost. The particle-particle interaction becomes so important that particles cannot be repelled from the progressing ice front.*

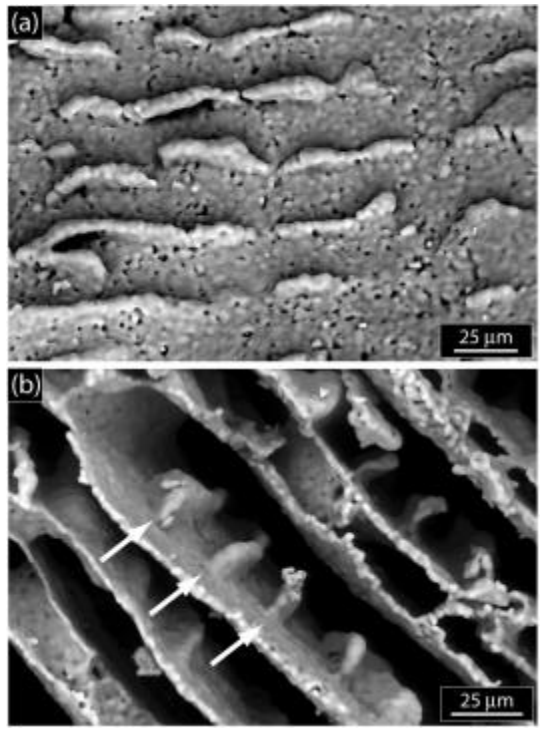

*Fig 12: Surface dendrites, oriented along the ice growth direction, cross section parallel (a) and perpendicular (b) to ice growth direction. These features might act as guidance pattern for cell growth, in addition to the already oriented porosity.*